\documentclass[%
 reprint,
 amsmath,amssymb,
 aps,
]{revtex4-2}

\usepackage{graphicx}% Include figure files
\usepackage{dcolumn}% Align table columns on decimal point
\usepackage{bm}

\usepackage{graphicx}

\usepackage[table]{xcolor}

\usepackage{multirow}

\usepackage{amsmath}

\DeclareUnicodeCharacter{2212}{-}

\draft % marks overfull lines with a black rule on the right

\begin{document}

%\preprint{}

\title[]{Suppressed weak anti-localization in topological insulator - antiferromagnetic insulator  (BiSb)$_2$Te$_3$ - MnF$_2$ thin film bilayers}
% Force line breaks with \\
\author{Ryan Van Haren}
\email{rvanhare@ucsc.edu}
    
\author{David Lederman}%

\affiliation{Physics Department, University of California Santa Cruz}

\date{\today}% It is always \today, today,

\begin{abstract}
Thin films of the topological insulator (BiSb)$_2$Te$_3$ oriented along the [0001] direction were grown via molecular beam epitaxy on substrates of Al$_2$O$_3$ (0001) and MgF$_2$ (110) single crystals, as well as on an epitaxial thin film of the antiferromagnetic insulator MnF$_2$ (110). Magnetoconductivity measurements of these samples showed close proximity of the Fermi level to the Dirac point and weak antilocalization at low temperature that was partially suppressed in the sample grown on the MnF$_2$ layer. The magnetoconductivity data were fit to a model that describes the quantum corrections to the conductivity for the Dirac surface state of a 3-dimensional topological insulator, from which values of the Fermi velocity and the phase coherence length of the surface state charge carriers were derived. The magnetoconductivity of the (BiSb)$_2$Te$_3$ - MnF$_2$ bilayer samples were fit to a model describing the crossover from weak antilocalization to weak localization due to magnetic doping. The results are consistent with the opening of an energy gap at the Dirac point in the (BiSb)$_2$Te$_3$ due to magnetic proximity interactions of the topological surface states with the antiferromagnetic MnF$_2$ insulator. 
\end{abstract}

\maketitle

\section{Introduction}

The interface between antiferromagnetic insulators and conductors with strong spin-orbit coupling has been shown to host interesting antiferromagnetic spintronic effects \cite{han_coherent_2023}. In Pt-MnF$_2$ bilayers specifically, generating a thermal gradient in the MnF$_2$ layer has the effect of inducing an electric voltage in the Pt overlayer through spin interactions via the spin Seebeck effect \cite{wu_antiferromagnetic_2016}. Spin-pumping has also been observed in in Cr$_2$O$_3$ and MnF$_2$ \cite{li_spin_2020,vaidya_subterahertz_2020}. In Pt-MnF$_2$ bilayers, magnon modes in the MnF$_2$ layer excited by subterahertz microwaves create an electric voltage in the Pt overlayer through spin interactions at the interface \cite{vaidya_subterahertz_2020}. The magnon modes of antiferromagnetic insulators like MnF$_2$ are especially intriguing for potential use in magnetoelectric devices, as the electronic transduction of these subterahertz and terahertz magnetic excitations could be used to fill the so-called terahertz gap \cite{sirtori_bridge_2002}. Coupling these modes to electric voltages, such as through spin-pumping into a Pt overlayer, is a necessary step to integrate them with contemporary semiconductor technology. From a fundamental point of view, MnF$_2$ is an Ising-like antiferromagnet with a rutile crystal structure and well-characterized magnetic interactions \cite{van_haren_emergent_2023}, making it a model system for studying the topological insulator - antiferromagnetic interface. 

Topological insulators (TIs) are promising magnetoelectronic materials when interfaced with an insulating antiferromagnet due to their exceptionally large surface spin currents generated by the spin-momentum locking in the topologically protected surface states \cite{mellnik_spin-transfer_2014, jamali_giant_2015}. Interface effects can be strong because the spin-momentum locked surface states are physically located on the edges or surfaces of two-dimensional (2D) and three-dimensional (3D) TIs, respectively, bringing the topologically protected conduction electrons in close spatial proximity to the magnetic moments of the antiferromagnetic layer. 

Short-ranged exchange interactions at a 3D TI-magnetic material interface could, in principle, affect the TI surface states and result in interesting quantum effects.  For example, TI - magnetic insulator interface can open an energy gap at the Dirac point in the normally gapless band structure of the TI surface states. These gapped surface states are expected to host unique quantum phenomena, perhaps most notably the quantum anomalous Hall effect. While TI - ferromagnetic or ferrimagnetic insulator bilayers are expected to open a surface state gap due to a nonzero magnetic moment normal to the film surface, it is unclear whether the same can be done with an antiferromagnet \cite{bhowmick_suppressed_2017}. In terms of magnetoelectronic devices, using an antiferromagnetic insulator can have the advantage that fringing fields from the antiferromagnet are negligible in comparison to those from a ferromagnet or ferrimagnet. Insulators have the additional advantage that the current is not shunted by the magnetic layer. 

When it comes to understanding the low-temperature magnetoconductivity of TI crystals and thin films, the Hikami-Larkin-Nagaoka (HLN) model is commonly used to fit to the weak localization (WL) and weak antilocalization (WAL) phenomena that TIs exhibit \cite{liu_crossover_2012, he_impurity_2011, nepal_disorder_2022, li_quantitative_2019, sultana_hikami-larkin-nagaoka_2018,shrestha_extremely_2017, le_thickness-dependent_2017}. Despite the pervasiveness of the HLN model in literature, it fails to account for several important features of the 3D-TI system. Specifically, the HLN model is derived for quasi-2D non-relativistic fermions, while the TI surface states are Dirac relativistic fermions, being described by the 2D Dirac Hamiltonian. In recent years, significant theoretical progress has been made in describing the behavior of the magnetoconductivity for systems with 2D Dirac states and strong spin-orbit coupling \cite{lu_finite-temperature_2014,adroguer_conductivity_2015}.

Here we study the electrical conductivity as a function of magnetic field and temperature of topological insulator (BiSb)$_2$Te$_3$ (BST) thin films grown on non-magnetic substrates of Al$_2$O$_3$ and MgF$_2$, and antiferromagnetic thin films of MnF$_2$ where the N\'eel vector lies in the plane of the film. These magnetoconductivity data are analyzed using a model of the quantum corrections to the magnetoconductivity of a 3D-TI that takes into account the presence of a 2D Dirac state and strong spin-orbit coupling. The results of this Dirac fermion model are compared with results from the quasi-2D HLN model. WAL is observed in all BST films but is suppressed in BST-MnF$_2$ thin film bilayers. Two explanations for the suppressed WAL behavior are proposed: an enhanced magnetic scattering from the local magnetic moments of the MnF$_2$ layer, and the opening of an energy gap at the Dirac point of the BST due to proximity with the antiferromagnetic MnF$_2$. To evaluate the former proposal, the results are analyzed in terms of the Dirac fermion model for massless fermions in a gapless surface state band structure. To evaluate the latter proposal, the magnetoconductivity data of the BST-MnF$_2$ bilayers are fit to a model describing the crossover from WAL to WL in a TI due to the opening of an energy gap at the Dirac point. The results of these analyses are discussed and further measurements to elucidate the situation are proposed.

\section{Methods}
\begin{figure}
    \centering
    \includegraphics{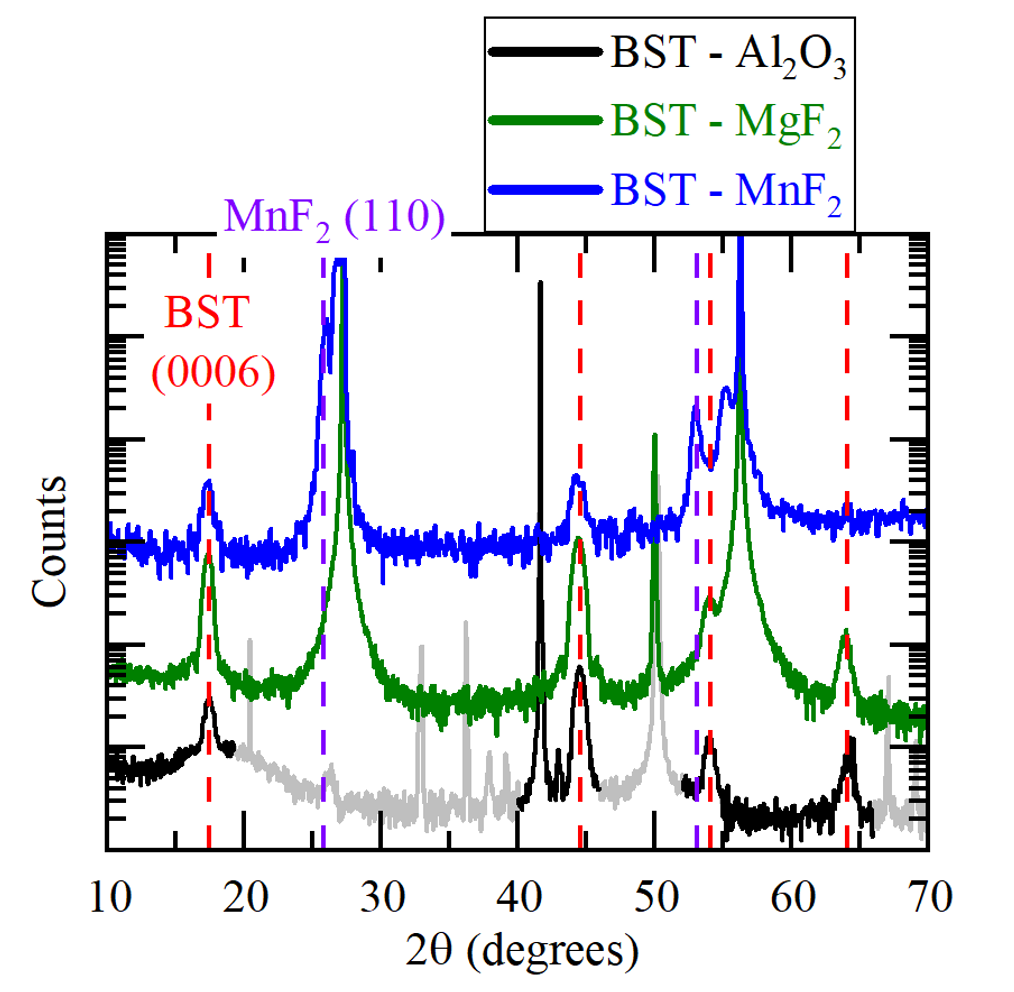}
    \caption{X-ray diffraction pattern of the three types of BST sample studied. At the bottom, shown in black, is 9 QL BST on Al$_2$O$_3$ (0001). XRD of this sample was performed after it was fabricated into a device and therefor has significant contaminants on the surface which manifest as anomalous XRD peaks, as shown in the semi-transparent black curve. In the middle, shown in green, is 9 QL BST on MgF$_2$ (110). At the top, shown in blue, is 9 QL BST grown on 30 nm MnF$_2$ (110) on a MgF$_2$ (110) substrate. The red dashed lines indicate the expected diffraction angles of the $\{0001\}$ planes of BST. The purple dashed lines indicate the expected diffraction angles of the $\{110\}$ planes of MnF$_2$.}
    \label{fig:BST-MnF2_XRD}
\end{figure}

Thin films of co-doped (Bi,Sb)$_2$Te$_3$ (BST) were grown via molecular beam epitaxy (MBE) on three types of substrate crystal: commercially purchased bulk crystals of Al$_2$O$_3$ (0001) and MgF$_2$ (110), and an MBE-grown epitaxial 30 nm thin film MnF$_2$. The MnF$_2$ thin film was grown using a graded buffer layer method to create a smooth and relaxed MnF$_2$ layer, as described in a previous work \cite{van_haren_emergent_2023}. The $(110)$ MnF$_2$ planes, whose surface normal is along the growth direction, have the magnetic easy axis (which is the axis of the N\'eel vector at low temperatures) of the crystal along the in-plane $[001]$ crystallographic direction (the $c$-axis). The BST films were grown to a thickness of 9 quintuple layers (QLs), or approximately 9~nm, and the stoichiometric ratio was approximately (Bi$_{0.35}$Sb$_{0.65}$)$_2$Te$_3$, as determined from the Bi/Sb flux ratio measured using a quartz crystal monitor, and x-ray reflectivity measurements of the film thickness. Before the growth of the BST layer, each substrate was first annealed at $T = 300 ^\circ$C for one hour. The first 3 QLs of BST were deposited in a Te flux surplus at a sample temperature of $T = 200 ^\circ$C. The flux was then shut off and the sample was heated to $T = 275 ^\circ$C, at which point 6 additional QLs of BST were deposited. After the films cooled to ambient temperature, but before they were removed from the vacuum growth chamber, 5 nm of amorphous MgF$_2$ was deposited to protect the BST surface from the atmosphere. The x-ray diffraction patterns of the three types of BST samples studied are shown in Fig.~\ref{fig:BST-MnF2_XRD}. The x-ray data are consistent with the orientation of the crystal structure as described above. 

Photolithography was used to develop a Hall bar pattern on the samples with a conducting channel $200\ \mu$m wide by $500\ \mu$m long. A wet etch process utilizing aqua regia was used to remove the unwanted film and complete the Hall bar. Electronic transport measurements were performed by wiring the Hall bar devices into a chip carrier using conducting silver paint and bare copper wire, mounting the chip carrier on a JANIS SuperVariTemp X-Gas Superconducting Magnet System, and performing DC voltage measurements with Keithley current sources and digital multimeters. The BST films host two independent conducting channels located at the top and bottom surfaces of the film. Because our BST films were very thin, the magnetoconductance measurements probed both channels simultaneously. The two BST-MnF$_2$ samples studied in this work are from two different Hall bar devices (labelled as \textit{e} and \textit{f}) fabricated from different pieces of the same original BST-MnF$_2$ thin film. The conducting channels are oriented with respect to the in-plane $c$-axis of the MnF$_2$, with the current density $\mathbf{J}$ in one channel parallel to the $c$-axis and the other perpendicular to the $c$-axis. Model fits to the experimental data were performed using the Levenberg-Marquardt non-linear least-squares algorithm included with OriginPro analysis software.

\begin{table*}
    \centering
    \begin{ruledtabular}
    \begin{tabular}{ccc}
 Sample& Carrier density ($10^{12}$ cm$^{-2}$)&Hall mobility (cm$^2$ V$^{-1}$ s$^{-1}$)\\
 \hline
         BST - Al$_2$O$_3$&  $-3.75 \pm 0.01$& $117.9 \pm 0.1$\\
         BST - MgF$_2$&  $-7.67 \pm 0.02$& $57.7 \pm 0.1$\\
         BST - MnF$_2$ (e) ($\mathbf{J} \parallel \mathbf{c}$)&  $-4.67 \pm 0.02$& $69.9 \pm 0.3$\\
         BST - MnF$_2$ (f) ($\mathbf{J} \perp \mathbf{c}$)&  $-4.55 \pm 0.05$& $82.7 \pm 0.9$\\
    \end{tabular}
    \end{ruledtabular}
    \caption{Carrier density and Hall mobility of BST samples calculated at $T = 2$ K. The negative values of the carrier densities indicate n-type carriers. $\mathbf{J} \parallel \mathbf{c}$ and $\mathbf{J} \perp \mathbf{c}$ indicate measurements performed with the current applied parallel and perpendicular to the MnF$_2$ in-plane $c$-axis, corresponding to the antiferromagnetic easy axis.}
    \label{tab:CarDen}
\end{table*}

\section{Results and discussion}

\begin{figure}
    \centering
    \includegraphics{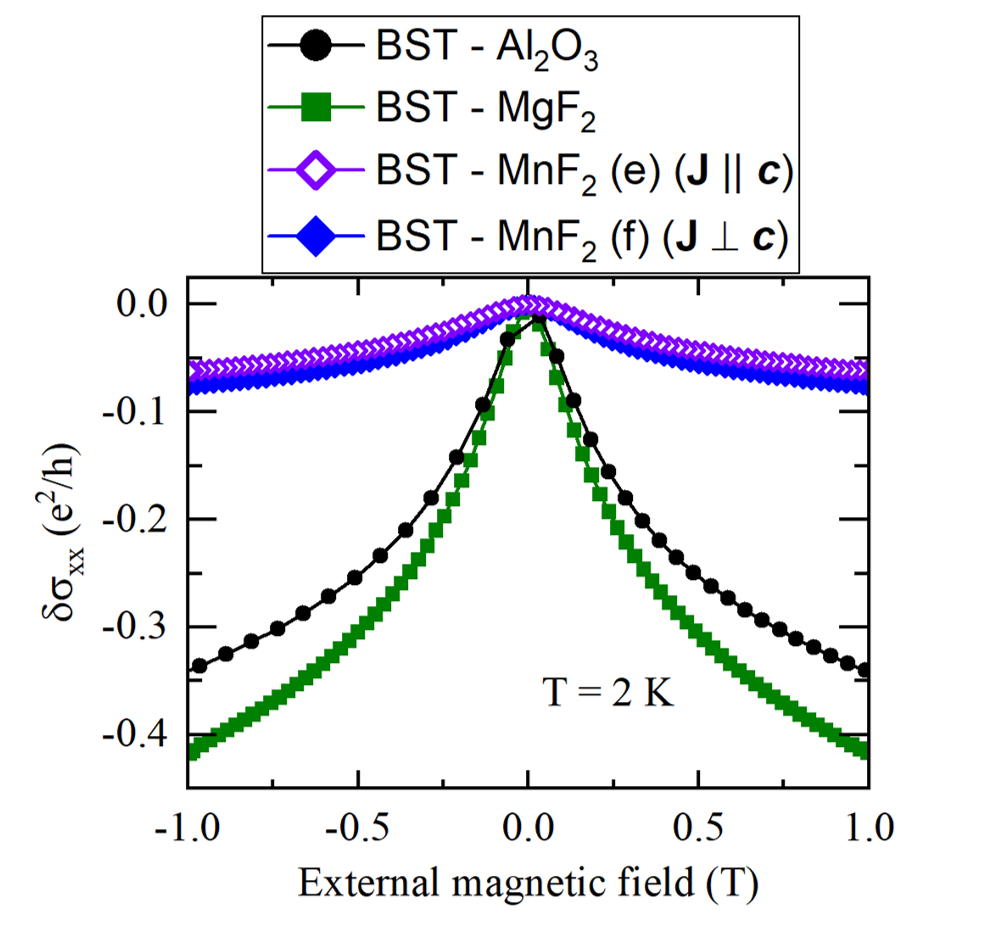}
    \caption{Electrical magnetoconductivity of the BST samples four samples studied, each exhibiting varying strengths of WAL.}
    \label{fig:RvH_1T}
\end{figure}

In Bi$_2$Te$_3$ and Sb$_2$Te$_3$ films, crystal defects tend to cause the Fermi level to move away from the Dirac point of the surface state in the bulk band gap and into the bulk bands. This results in the carrier density of the material increasing and electronic conduction being shunted through the bulk states of the TI, where the topological properties of the surface states do not apply. By co-doping n-type Bi$_2$Te$_3$ with p-type Sb$_2$Te$_3$ at the appropriate level, the Fermi level can be engineered back into the bulk band gap to enhance conduction through the surface states \cite{zhang_band_2011}. The BST films studied here are all n-type, with 2D carrier densities near the crossover from n-type to p-type, indicating that the Fermi level is very near the Dirac point and 2D surface state transport is significant. The carrier densities and Hall mobilities of each of the samples at $T = 2$ K are given in table~\ref{tab:CarDen}, with the carrier densities suggesting that the Fermi level lies between 0.1 and 0.2 eV above the Dirac point \cite{zhang_band_2011}. Each of these samples exhibits the positive magnetoconductivity cusp at low temperature and small external magnetic fields associated with WAL, as shown in Fig.~\ref{fig:RvH_1T}.

In ordinary metals, the electrical conductivity decreases at low temperature due to electron-electron interactions \cite{altshuler_interaction_1980}. In 2D systems, such as the surface states of a 3D TI, WAL and WL effects may also appear at low temperature due to the Berry phase that the electron wavefunction acquires as it scatters \cite{moessner_topological_2021}. In the case of the BST films studied here, only WAL is present. Measuring the conductivity of these 2D systems as a function of temperature can yield insight into the elastic scattering of the charge carriers that will be useful for further analysis. The correction to the conductivity, resulting from electron-electron interactions and WAL, as a function of temperature is expressed as
\begin{eqnarray}\label{Eq:AAL_MCvT}
        \delta \sigma(T) = \sigma(T) - \sigma(T_0) = \frac{e^2}{2 \pi^2 \hbar}\alpha p \ln\left(\frac{T}{T_0}\right) \\ +  
        \frac{e^2}{4 \pi^2 \hbar} (2 - 2F) \ln\left(\frac{k_B T \tau_e}{\hbar}\right), \nonumber
\end{eqnarray}
where $\alpha$ is a dimensionless parameter that indicates the presence of weak localization or antilocalization, defined in the same way as in the HLN model,  $p$ is another dimensionless parameter that depends on the dominant collision mechanism, $F$ is a scaling factor for the Hartree term, $\tau_e$ is the electron relaxation time due to elastic scattering processes, and $T_0$ is a characteristic temperature defined as $T_0 = \hbar/k_B \tau_e$  \cite{lee_disordered_1985, jana_evidence_2021, altshuler_magnetoresistance_1980, kumar_electron-electron_2020}.

\begin{figure}
    \centering
    \includegraphics{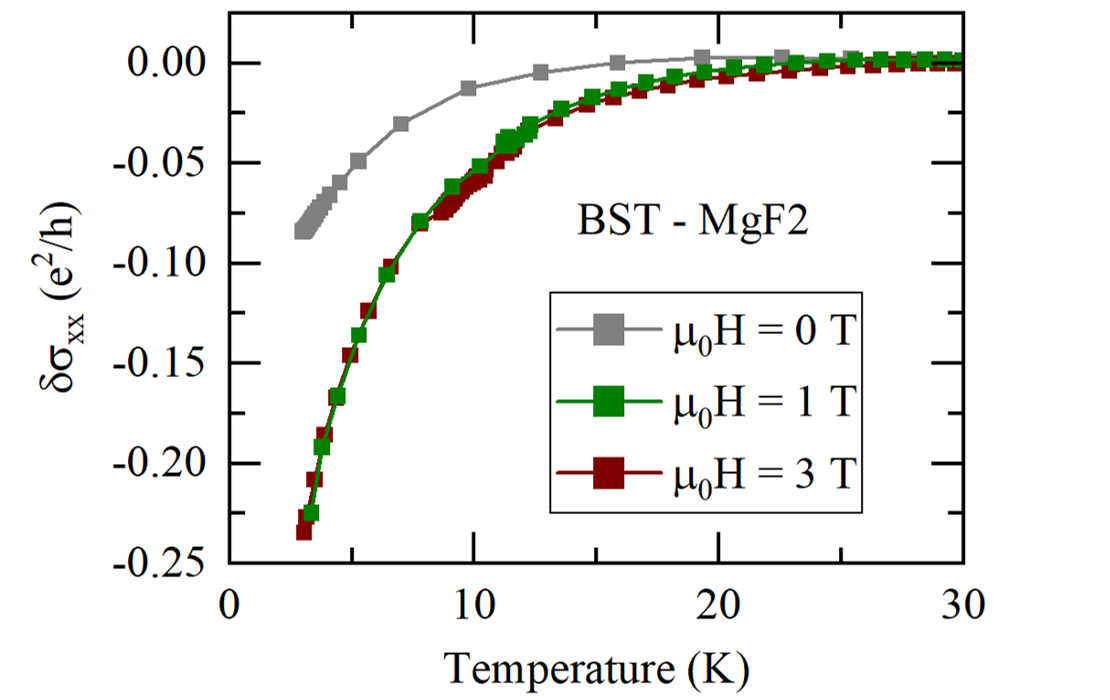}
    \caption{Electrical magnetoconductivity of the BST - MgF$_2$ sample measured as a function of temperature in several different external magnetic field strengths.}
    \label{fig:RvT_highB}
\end{figure}

When measuring the magnetoconductivity as a function of temperature in TI films, WAL has the effect of increasing the conductivity at low temperature, competing with electron-electron interactions that tend to decrease the conductivity. At sufficiently large external magnetic fields applied perpendicular to the 2D conducting layer, the quantum interference term $\alpha p$ in Eq.~\ref{Eq:AAL_MCvT} responsible for WAL is suppressed. The strength of the perpendicular magnetic field necessary to suppress the WAL depends on the phase coherence length of the charge carriers, and therefore varies between materials and samples \cite{altshuler_magnetoresistance_1981}. In the films studied here, an external field of $\mu_0 H = 1.0$ T is sufficient to fully suppress WAL, as shown in Fig.~\ref{fig:RvT_highB}, where the decrease in the electrical conductivity is smallest for zero external magnetic field due to WAL and the decrease in conductivity saturates by $\mu_0 H = 1$ T, an effect that has been observed in other materials \cite{jana_evidence_2021}.

\begin{figure*}
    \centering
    \includegraphics{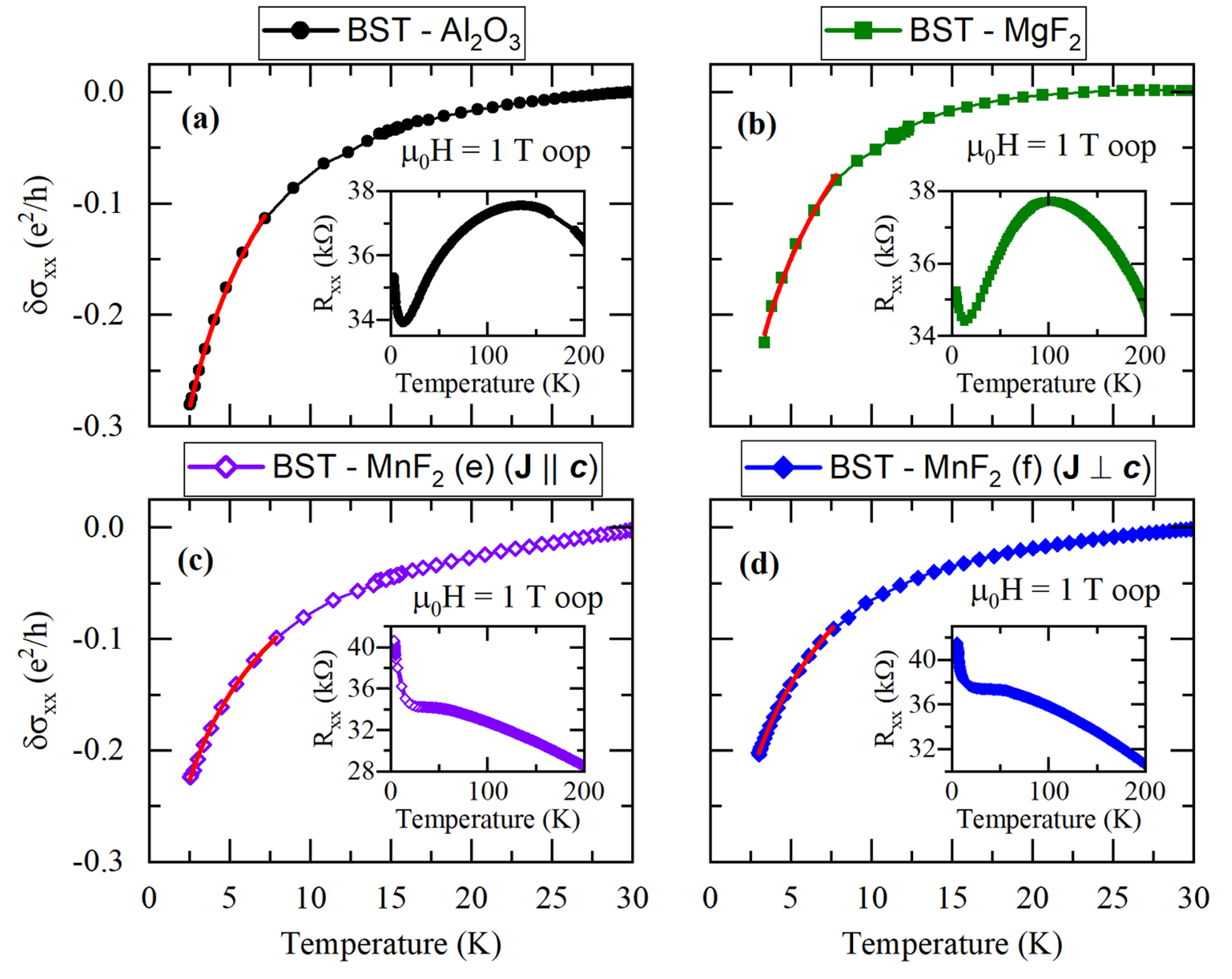}
    \caption{Electrical magnetoconductivity of several BST films as a function of temperature in an external magnetic field of $\mu_0 H = 1$ T, applied normal to the film surface. (a) shows BST grown on Al$_2$O$_3$ (0001). (b) shows BST grown on MgF$_2$ (110). (c) shows BST grown on thin film MnF$_2$ (110), with the current applied parallel to the $c$-axis of the MnF$_2$ crystal. (d) also shows BST grown on thin film MnF$_2$ (110), but with the current applied perpendicular to the $c$-axis of the MnF$_2$ crystal. Insets show the longitudinal resistance of the samples over a larger range of temperature.}
    \label{fig:MCvT}
\end{figure*}

In the case of a large external magnetic field, the 2D conducting layer is in the unitary case where $\alpha = 0$ and the first term in Eq.~\ref{Eq:AAL_MCvT} goes to zero \cite{hikami_spin-orbit_1980}. The temperature dependent magnetoconductivity for a system with two conducting channels is then expressed as
\begin{equation}\label{Eq:AAL_MCvT_2}
    \delta \sigma (B=1\text{ T},T)= 2 \times \frac{e^2}{4 \pi^2 \hbar} (2 - 2F) \ln\left(\frac{k_B T \tau_e}{\hbar}\right).
\end{equation}
The red curves in Fig.~\ref{fig:MCvT} represent non-linear least-square fits of the magnetoconductivity to Eq.~\ref{Eq:AAL_MCvT_2}. The values for $F$ and the elastic scattering time $\tau_e$ obtained from the fits are presented in table~\ref{tab:AAL_MCvT}.

\begin{table}
    \centering
    \begin{ruledtabular}
    \begin{tabular}{ccc}
         Sample&  $F$& $\tau_e$ (ps)\\
         \hline
         BST - Al$_2$O$_3$&  $0.74 \pm 0.01 $& $0.086 \pm 0.001$\\
         BST - MgF$_2$&  $0.73 \pm 0.01 $& $0.100 \pm 0.004$\\
 BST - MnF$_2$ (e) ($\textbf{J} \parallel \textbf{c}$)& $0.82 \pm 0.01 $&$0.064 \pm 0.001$\\
 BST - MnF$_2$ (f) ($\textbf{J} \perp \textbf{c}$)& $0.80 \pm 0.01 $&$0.076 \pm 0.001$\\
    \end{tabular}
    \caption{Values derived from fitting Eq.~\ref{Eq:AAL_MCvT_2} to the temperature-dependent magnetoconductivity data shown in Fig.~\ref{fig:MCvT}.}
    \label{tab:AAL_MCvT}
\end{ruledtabular}
\end{table}

\begin{figure}
    \centering
    \includegraphics{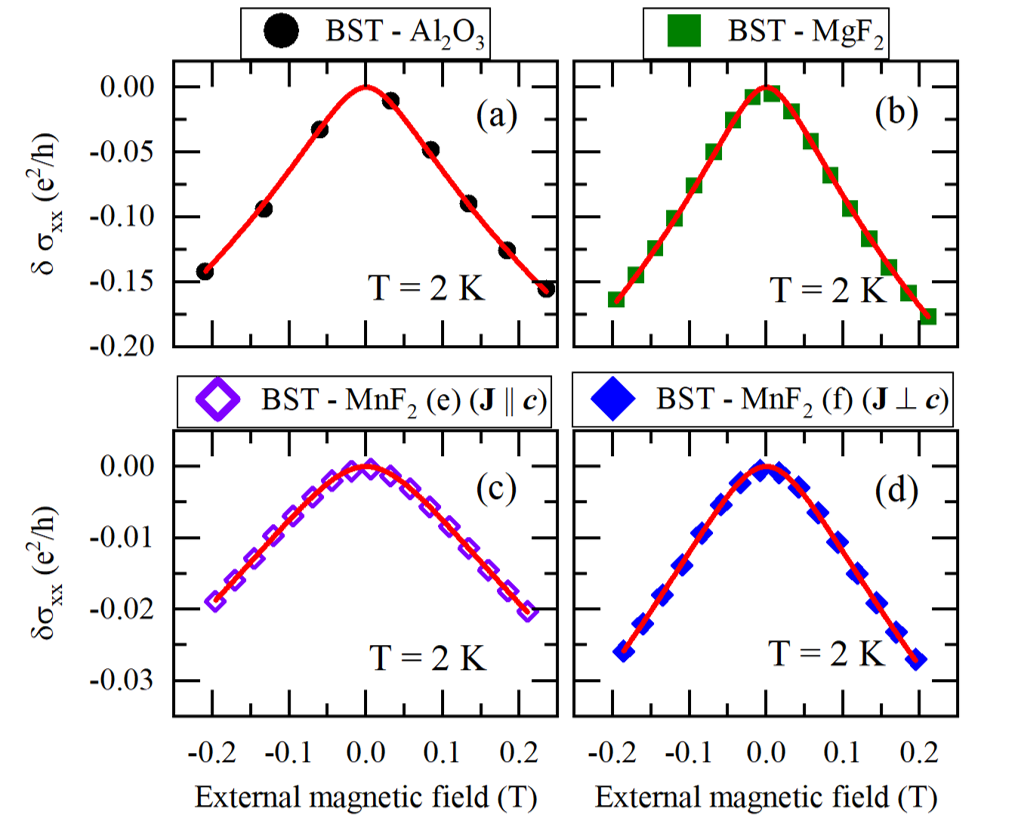}
    \caption{Electrical magnetoconductivity of several BST films as a function of the applied magnetic field at $T = 2$ K. Fits to the HLN model given in Eq.~\ref{eq:HLN} are shown in red. Derived parameters from the fits are given in table~\ref{tab:RvH_HLN}.}
    \label{fig:RvH_HLN}
\end{figure}

The HLN model used to fit to the magnetoconductivity of the TI films at a fixed temperature is expressed as
\begin{equation}
\label{eq:HLN}
    \delta \sigma(B) = -\frac{\alpha e^2}{\pi h} \left [ \ln\left(\frac{\hbar}{4 e l_\phi^2 B}\right) - \psi \left ( \frac{1}{2} + \frac{\hbar}{4 e l_\phi^2 B} \right ) \right ],
\end{equation}
where $\psi$ is the digamma function, $l_\phi$ is the phase coherence length, and $\alpha$ is a dimensionless parameter that describes the type of localization. The parameter $\alpha$ nominally takes on one of three possible values: $\alpha = 1$ is known as the orthogonal case that occurs when there is no SOC and no magnetic scattering, weak localization occurs in the orthogonal case; $\alpha = 0$ is the unitary case where there are no localization effects and there is no quantum localization correction to the magnetoconductivity; and $\alpha = - 1/2$ is the symplectic case and corresponds to strong SOC and no magnetic scattering. WAL occurs only in the symplectic case \cite{hikami_spin-orbit_1980}. Fits of the HLN model to the magnetoconductivity of the same BST films presented above are shown in Fig.~\ref{fig:RvH_HLN}. The derived parameters from the HLN model fits are given in table~\ref{tab:RvH_HLN}

\begin{table}
    \centering
    \begin{ruledtabular}
    \begin{tabular}{ccc}
         Sample & $\alpha$ & $l_{\phi,\text{HLN}}$ (nm)\\
         \hline
         BST - Al$_2$O$_3$& $-0.71 \pm 0.05$& $79 \pm 3$\\
         BST - MgF$_2$ & $-0.77 \pm 0.03$& $86 \pm 2$\\
 BST - MnF$_2$ (e) ($\textbf{J} \parallel \textbf{c}$)& $-0.19 \pm 0.01$&$59 \pm 1$\\
 BST - MnF$_2$ (f) ($\textbf{J} \perp \textbf{c}$)& $-0.20 \pm 0.01$ & $67 \pm 1$\\
    \end{tabular}
    \caption{Values derived from fitting the HLN model Eq.~\ref{eq:HLN} to the magnetoconductivity data shown in Fig.~\ref{fig:RvH_HLN}(b).}
    \label{tab:RvH_HLN}
\end{ruledtabular}
\end{table}

While the HLN model does not capture the full physical behavior of the Dirac surface states of the TI, it still provides useful insight into the localization behavior present in the films through the parameter $\alpha$. In the case of TIs, it is expected that the top and bottom surfaces of the layer can act as independent 2D conducting channels, each contributing an additive factor of $\alpha = -0.5$, if both channels exhibit WAL. This is why some TI films can have $\alpha$ values more negative than $-0.5$ ~\cite{wang_thickness-dependent_2016}. According to the results shown in Table~\ref{tab:RvH_HLN}, the BST films on the non-magnetic substrates  yield an $\alpha$ value of approximately $-0.75$, suggesting the presence of two independent conducting 2D channels, each exhibiting some degree of WAL. In contrast, the BST film on antiferromagnetic MnF$_2$ shows that $\alpha = -0.2$. In the framework of the HLN model, $\alpha$ will approach and eventually reach a value of zero as the strength of the magnetic scattering in the quasi-2D conducting channel increases. This has been observed experimentally in ferromagnetic TI films doped with transition metal ions, where the introduction of magnetic scatterers suppress WAL and cause $\alpha$ to approach zero, or even cause a crossover from WAL to weak localization \cite{liu_crossover_2012, van_haren_surface_2023}. In these BST-MnF$_2$ bilayers, the only source of magnetic scattering is from the TI interface with the antiferromagnetic MnF$_2$ layer, suggesting the presence of some magnetoelectronic interaction at the interface.

\begin{figure}
    \centering
    \includegraphics{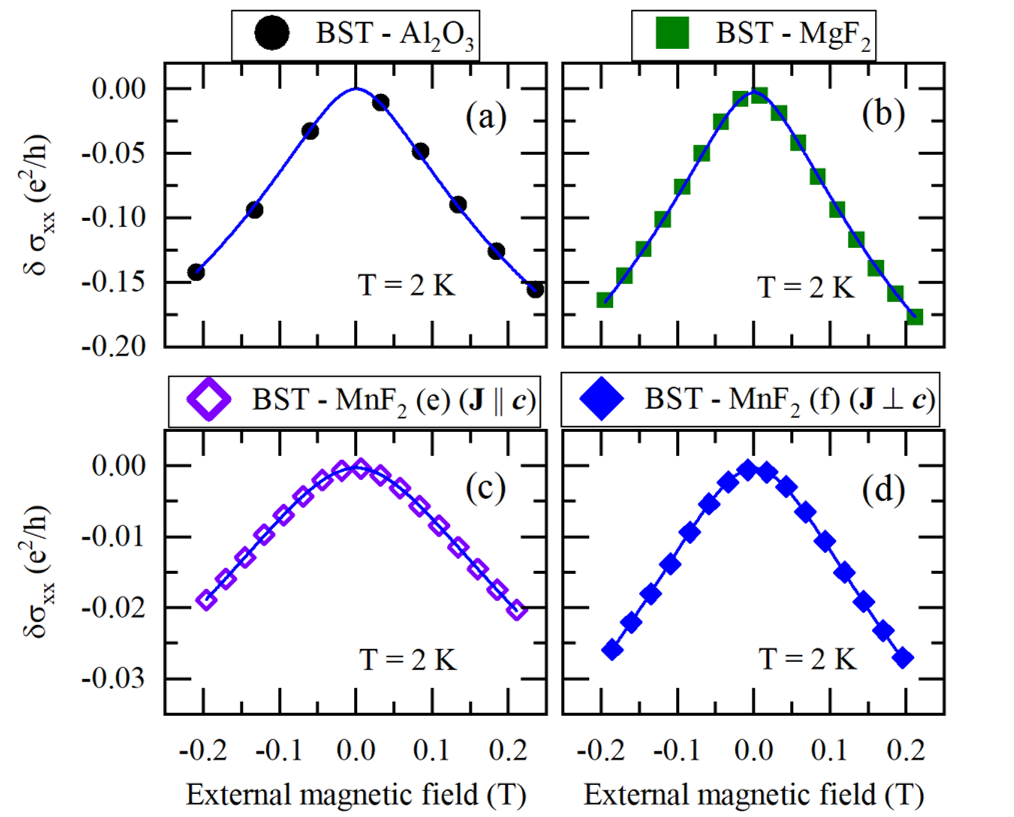}
    \caption{Electrical magnetoconductivity of several BST films as a function of the applied magnetic field at $T = 2$ K. Fits to the Dirac fermion model, presented in Eq.~\ref{Eq:DiracWAL}, are shown in blue. The derived parameters from these fits are given in table~\ref{tab:RvH_Dirac_2}.}
    \label{fig:RvH_Dirac}
\end{figure}

In contrast with the quasi-2D HLN model, the Dirac fermion model developed by Adroguer et al. takes into account the 2D Dirac nature of the TI surface state \cite{adroguer_conductivity_2015}. In this model, the magnetoconductivity correction for each conducting channel has contributions from three different propagation modes, and in the case of two channels, is expressed as
\begin{equation} \label{Eq:DiracWAL}
        \delta \sigma (B) = 2 \frac{e^2}{ \pi h} \sum_{i=1}^3 a_i w_i g_i(B),
\end{equation}
where 
\begin{eqnarray}
         a_1 w_1 g_1 (B) = \left( \frac{1}{2} - \tilde{\lambda} + \frac{3}{2} \tilde{\lambda}^2 \right) \times \\
        \left[ \psi \left( \frac{1}{2} + \frac{\hbar (1 + \frac{1}{2} \tilde{\lambda}^2)}{4 e \nu_F^2 \tau_e^2 B} \right) - \psi \left( \frac{1}{2} + \frac{\hbar (1 + \frac{1}{2} \tilde{\lambda}^2)}{4 e \nu_F^2 \tau_e \tau_\phi B} \right) \right], \nonumber
\end{eqnarray}
\begin{eqnarray}
     a_2 w_2 g_2 (B) =  \left( \tilde{\lambda} - 2 \tilde{\lambda}^2 \right)  \times \\
        \left[ \psi \left( \frac{1}{2} + \frac{\hbar \tilde{\lambda}}{8 e \nu_F^2 \tau_e^2 B} \right) - \psi \left( \frac{1}{2} + \frac{\hbar \tilde{\lambda}}{8 e \nu_F^2 \tau_e \tau_\phi B} \right) \right], \nonumber
\end{eqnarray}
\begin{eqnarray}
     a_3 w_3 g_3 (B) =  \frac{1}{2} \tilde{\lambda}^2  \times \\
        \left[ \psi \left( \frac{1}{2} + \frac{\hbar \tilde{\lambda}^2}{8 e \nu_F^2 \tau_e^2 B} \right) - \psi \left( \frac{1}{2} + \frac{\hbar \tilde{\lambda}^2}{8 e \nu_F^2 \tau_e \tau_\phi B} \right) \right].  \nonumber
\end{eqnarray}
Here $\tau_\phi$ and $\tau_e$ represent the spin-orbit and elastic scattering times respectively, $\tilde{\lambda}$ represents the relative strength of the scalar and spin-orbit coupled disorder, and $\nu_F$ represents the Fermi velocity. In the Dirac fermion model, it is necessary that $\tilde{\lambda} << 1$, meaning that the spin orbit scattering time $\tau_\phi$ is much larger than the elastic scattering time $\tau_e$. The Dirac fermion model predicts that a TI will always be in the symplectic symmetry class where WAL is present. Note that a factor of $\ln(\tau_\phi/\tau_e)$ inside and outside the sum are cancelled out in the above expressions relative to the original paper~\cite{adroguer_conductivity_2015}. 
Figure~\ref{fig:RvH_Dirac} shows the low temperature magnetoconductivity of the same four BST samples fit to the Dirac fermion model presented in Eq.~\ref{Eq:DiracWAL}, with a background offset of $C = -(1/2\pi)\ln(\tau_\phi/\tau_e)$ per conducting channel added to align the fits with the origin $\sigma(0) = 0$. There is a high amount of dependency between the fitting parameters $\tau_\phi$ and $\tau_e$ in the Dirac fermion model that make good fits difficult to achieve \cite{zhou_quantum_2020}. This issue is ameliorated by making the assumption that the elastic scattering time $\tau_e$ derived from fitting the conductivity data, shown in Fig.~\ref{fig:MCvT} to Eq.~\ref{Eq:AAL_MCvT_2} and given in table~\ref{tab:AAL_MCvT}, is an accurate description of the elastic scattering time in the Dirac fermion model. By making this assumption, $\tau_e$ can be held constant to the previously derived value and the dependency between the parameters is greatly reduced. 

The derived parameters from the Dirac fermion model fit to the data are given in table~\ref{tab:RvH_Dirac_2}.  The spin coherence length from the Dirac fermion model, $l_{\phi,\text{DF}}$, is derived according to the expression $l_{\phi,\text{DF}} = \tau_\phi \nu_F$. The spin coherence lengths of the BST films grown on non-magnetic substrates derived from the Dirac fermion model are about twice as large as the same values derived from the HLN model, but in the case of the BST-MnF$_2$ samples, the values are very similar between the two models. This could suggest that the HLN model is less accurate in describing the transport behavior of a pure TI where 2D Dirac transport dominates, but that this accuracy improves when magnetic scattering is strengthened and the topological protection of the Dirac surface states is weakened, as in the BST-MnF$_2$ samples. The derived Fermi velocities for the BST films studied here fall in the range between $300$ km/s and $600$ km/s, consistent with the values commonly found from ARPES measurements \cite{chen_experimental_2009, zhang_band_2011}. This result demonstrates the utility of the Dirac fermion model as implemented here, as it enables the determination of the Fermi velocity from magnetoconductivity measurements alone. 

\begin{table*}
\begin{ruledtabular}

    \centering
    \begin{tabular}{ccccc}
         Sample &  $\nu_F$ (km/s) &  $\tilde{\lambda}$ &  $\tau_\phi$ (ps) & $l_{\phi,\text{DF}}$ (nm) \\
         \hline
         BST - Al$_2$O$_3$&  $406 \pm 42$&  $0 \pm 0.003$&  $0.36 \pm 0.05$& $149 \pm 7$\\
         BST - MgF$_2$&  $311 \pm 15 $&  $0.003 \pm 0.001$&  $0.62 \pm 0.05$& $191 \pm 6$\\
         BST - MnF$_2$ (e) ($\textbf{J} \parallel \textbf{c}$)&  $607 \pm 6$&  $0.0016 \pm 0.0002 $&  $0.087 \pm 0.001 $& $52.7 \pm 0.2$\\
         BST - MnF$_2$ (f) ($\textbf{J} \perp \textbf{c}$)&  $572 \pm 8$&  $0.0018 \pm 0.0004$&  $0.107 \pm 0.001$& $61.3 \pm 0.3$\\
    \end{tabular}
    \caption{Values derived from fitting the Dirac fermion model Eq.~\ref{Eq:DiracWAL} to the magnetoconductivity data shown in Fig.~\ref{fig:RvH_Dirac}.}
    \label{tab:RvH_Dirac_2}
    \end{ruledtabular}
\end{table*}

In the specific case of the BST-MnF$_2$ bilayers, the suppression of WAL relative to the BST films on nonmagnetic substrates are accompanied by a smaller phase coherence length, evident in both the HLN and Dirac fermion models. This result is consistent with the interpretation that enhanced magnetic scattering at the antiferromagnetic MnF$_2$ interface is the source of the suppressed WAL, as magnetic scattering sites at the interface will cause the electron phase lose coherence more quickly when compared to a sample without magnetic scatterers. 

\begin{figure}
    \centering
    \includegraphics{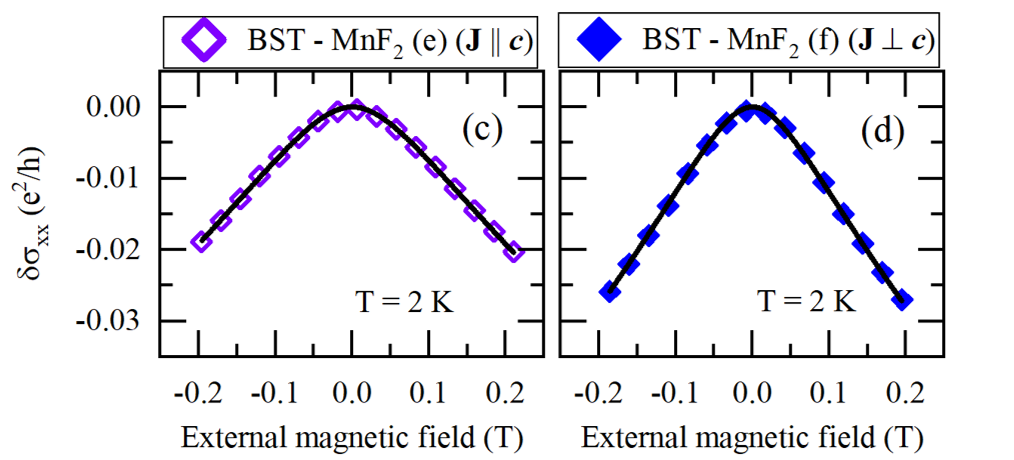}
    \caption{Electrical magnetoconductivity of the BST-MnF$_2$ bilayers at $T = 2$ K fit to the massive fermion model given in Eq.~\ref{Eq:massivefermions}. Derived parameters from the fit are given in table~\ref{tab:massive_fermions}.}
    \label{fig:RvH_MassFermions}
\end{figure}

Another possible explanation for the suppression of WAL is that a magnetically induced energy gap $\Delta$ is created at the Dirac point of the TI surface state. In the presence of magnetic dopants, the ordinary WAL behavior of a TI can be suppressed and eventually supplanted by weak localization as the size of the gap increases and the surface state fermions become massive \cite{lu_competition_2011}. The model used to describe this phenomenon in the case of two conducting channels is expressed as \cite{lu_competition_2011,zheng_weak_2016}
\begin{equation}\label{Eq:massivefermions}
    \delta \sigma (B) = 2 \sum_{i = 0,1} \frac{\alpha_i e^2}{\pi h} \left[ \psi \left( \frac{1}{2} +\frac{l^2_B}{l^2_{\phi i}} \right) - \ln\left( \frac{l^2_B}{l^2_{\phi i}} \right) \right],
\end{equation}
where
\begin{equation}
\begin{split}
    l^2_B &= \frac{\hbar}{4 e B}, \\
    l^2_{\phi i} &= \left( \frac{1}{l^2_\phi} + \frac{1}{l^2_i} \right), \\
    l^2_0 &= l^2_e a^4 (a^4 + b^4 - a^2b^2)/[b^4(a^2 - b^2)^2], \\
    l^2_1 &= l^2_e(a^4 + b^4)^2/[a^2b^2(a^2 - b^2)^2], \\
     \alpha_0 &= (a^4 + b^4)(a^2 - b^2)^2/[2(a^4 + b^4 - a^2b^2)^2], \\
    \alpha_1 &= -a^4b^4/[(a^4 + b^4)(a^4 + b^4 -a^2b^2)],
\end{split}
\end{equation}
and $l_e$ is the mean free path, $a = \cos(\theta/2)$, and $b = \sin(\theta/2)$. The relationship between the Fermi energy and the Dirac point gap is given by $\cos(\theta) = \Delta/2E_F$, where $E_F$ is the Fermi energy. Data from the BST-MnF$_2$ bilayer samples were fit using this model as shown in Fig.~\ref{fig:RvH_MassFermions}. The relative size of the energy gap $\Delta/2E_F$  and the phase coherence length $l_\phi$ derived from the fit are given in table~\ref{tab:massive_fermions}. For both MnF$_2$ samples,   $\Delta/2E_F \approx 0.28$, suggesting the development of a small energy gap at the Dirac point of about half the magnitude of the Fermi energy. The magnitude of this gap is consistent with previous results in TIs using this model, including BST- (insulating ferrimagnetic) BaFe$_{12}$O$_{19}$ bilayers~\cite{zheng_weak_2016} and BST -(metallic ferromagnetic) Fe$_7$Se$_8$  bilayers~\cite{jarach_carrier_2021}. The relatively small value of $\delta \sigma_{xx}$ and rounding of the WAL cusp are consistent with what is expected qualitatively from the model as the energy gap $\Delta$ increases from zero \cite{lu_competition_2011}. 

Both the nonzero value of $\Delta/2E_F$ and the shortened phase coherence length relative to BST samples on non-magnetic substrates are consistent with the development of a surface state energy gap due to the antiferromagnetic MnF$_2$ interface. However, the opening of an energy gap usually requires a net magnetic moment normal to the film surface and direction of charge transport, either through magnetic doping or proximity to a ferromagnetic or ferrimagnetic material. The thin film MnF$_2$ used in these bilayers is (110) orientation, and not expected to contribute any net magnetization normal to the film surface \cite{van_haren_emergent_2023}. Note, however, that the exchange coupling to the antiferromagnet could result in an effective field perpendicular to the surface via, for example, a Dzyaloshinskii-Moriya interaction, depending on the local interface crystal symmetry.  Further measurements will need to be performed to determine conclusively if a gap is being opened in the BST. Performing ARPES, for example, should show the shape and structure of the surface state band, from which the presence of a gap would be readily observed. Alternatively, electronic transport measurements performed at temperatures below $T = 2$ K in samples with the Fermi level very close to the Dirac point could reveal evidence of the quantum anomalous Hall effect, which would be compelling evidence of a surface state gap. 

Finally, we note that the results for $l_\phi$ are slightly different when the current is applied parallel and perpendicular to the N\'eel vector of the AF. This could indicate that the magnetic interaction between the antiferromagnetic moments and the TI surface states depends on the relative orientation of the spins of the surface states with respect to the N\'eel vector of the AF as a result of the spin-momentum locking of the surface states. When $\bf{J}\parallel \bf{c}$, conduction electrons have a spin perpendicular to the N\'eel vector of the MnF$_2$, while the spins are parallel to the N\'eel vector when $\bf{J}\perp\bf{c}$. This result would be expected from an anisotropic exchange interaction between the surface states and the magnetic order of the AF at the interface.

\begin{table}
\begin{ruledtabular}

    \centering
    \begin{tabular}{ccc}
         Sample&  $l_\phi$ (nm)& $\Delta/2E_F$ \\
         \hline
         BST - MnF$_2$ (e) ($\textbf{J} \parallel \textbf{c}$)&  $61 \pm 1$&  $0.282 \pm 0.001$\\
         BST - MnF$_2$ (f) ($\textbf{J} \perp \textbf{c}$)&  $70 \pm 1$&  $0.279 \pm 0.001$\\
    \end{tabular}
    \caption{Results of fits of the model described in Eq.~\ref{Eq:massivefermions} to the data shown in Fig.~\ref{fig:RvH_MassFermions}. The mean free path is set to $l_e = 40$ nm for these fits, consistent with results from the Dirac fermion model.}
    \label{tab:massive_fermions}
\end{ruledtabular}
\end{table}

\section{Conclusions}

In this work, we have shown how the WAL behavior of a TI can be fit to a theoretical model that takes into account the Dirac nature and strong spin orbit coupling of the TI surface state and can be used to derive the Fermi velocity of the charge carriers and the spin-orbit scattering time. Fits of the Dirac fermion model are compared with fits of the HLN model to experimental magnetoconductivity data of BST films grown on non-magnetic substrates and on antiferromagnetic MnF$_2$ thin films. WAL is suppressed in the BST-MnF$_2$ bilayers, suggesting the presence of some magnetoelectric interaction at the interface of the two layers. Evidence for enhanced magnetic scattering due to magnetic scattering sites is provided by the shortened phase coherence lengths of the BST-MnF$_2$ bilayers. Fitting the magnetoconductivity of the BST-MnF$_2$ bilayers to a model describing the cross over from WAL to WL due to magnetic doping suggests that an energy gap may be forming at the Dirac point of the surface state due to proximity with the antiferromagnetic MnF$_2$ layer. We also find evidence for an anisotropic exchange interaction between the TI surface states and the AF spins. The techniques and results presented in this work should provide a starting point for further analysis of TI films and bilayers using the Dirac fermion model and serve to highlight potentially interesting magnetoelectric interactions in TI-antiferromagnetic insulator bilayers. 

\bibliography{BST-MnF2}% Produces the bibliography via BibTeX.

\end{document}